\documentclass[
 superscriptaddress,
 reprint,
 amsmath, amssymb,
 aps, prd, preprintnumbers, nofootinbib, floatfix
]{revtex4-1}

\usepackage{graphicx}
\usepackage{dcolumn}
\usepackage{bm}
\usepackage{color}
\usepackage{amssymb}
\usepackage{amsmath}
\usepackage{braket}
\usepackage[colorlinks=true,allcolors=Blue,pdfusetitle]{hyperref}
\usepackage{rotating}
\usepackage{multirow}
\usepackage{graphicx}
\usepackage[dvipsnames]{xcolor}
\usepackage{svg}
\usepackage{esdiff}
\definecolor{Green}{RGB}{0, 128, 0}
\newcommand{\orcid}[1]{\href{https://orcid.org/#1}{\includegraphics[width=10pt]{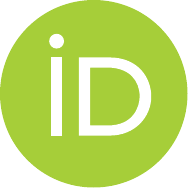}}}

\begin{document}
\preprint{N3AS-22-023}
\preprint{INT-PUB-22-051}
\title{Entanglement in three-flavor collective neutrino oscillations}

\author{Pooja Siwach \orcid{0000-0001-6186-0555}}
\email{psiwach@physics.wisc.edu}
\affiliation{
Department of Physics, University of Wisconsin--Madison,
Madison, Wisconsin 53706, USA}

\author{Anna M. Suliga \orcid{0000-0002-8354-012X}}
\email{asuliga@berkeley.edu}
\affiliation{
Department of Physics, University of California Berkeley, Berkeley, California 94720, USA}
\affiliation{
Department of Physics, University of Wisconsin--Madison,
Madison, Wisconsin 53706, USA}

\author{A. Baha Balantekin \orcid{0000-0002-2999-0111}}
\email{baha@physics.wisc.edu}
\affiliation{
Department of Physics, University of Wisconsin--Madison,
Madison, Wisconsin 53706, USA}

\date{November 14, 2022}


\begin{abstract}
Extreme conditions present in the interiors of the core-collapse supernovae make neutrino-neutrino interactions not only feasible but dominant in specific regions, leading to the non-linear evolution of the neutrino flavor. Results obtained when such collective neutrino oscillations are treated in the mean-field approximation deviate from the results using the many-body picture because of the ignored quantum correlations. We present the first three flavor many-body calculations of the collective neutrino oscillations. The entanglement is quantified in terms of the entanglement entropy and the components of the polarization vector. We propose a qualitative measure of entanglement in terms of flavor-lepton number conserved quantities. We find that in the cases considered in the present work, the entanglement can be underestimated in two flavor approximation. The dependence of the entanglement on mass ordering is also investigated. We also explore the mixing of mass eigenstates in different mass orderings.
\end{abstract}

\maketitle



\section{Introduction}
\label{sec:introduction}

One of the most complex yet-to-be-understood phenomena occurs during the last seconds of the massive star's evolution, right before it forms a compact object. The core of the star becomes too heavy to support itself against gravitational force triggering the inevitable gravitational collapse. However, once the nuclear saturation density is reached, the matter in the inner cannot be compressed anymore, and the core bounces back, releasing an energetic shock wave, which may trigger the supernova explosion. The hydrodynamical simulations show that the shock does not travel far; it stalls inside the star after only approximately tens of milliseconds (see, e.g., Refs.~\cite{Janka:2016fox, Horiuchi:2018ofe, Burrows:2020qrp} for recent reviews). It is due to the fact that the shock loses a significant fraction of its energy to the photodissociation of the nuclei and ram pressure of the still-infalling on core matter~\cite{Bethe:1979zd, Fuller:1981mv, 1984NuPhA.429..527C}. Therefore, for a star to explode successfully, a mechanism must exist that rejuvenates the stalled shock wave.

The delayed neutrino heating mechanism~\cite{Colgate:1966ax, Bethe:1985sox}, in which neutrinos oozing from the proto-neutron star revive the stalled shock wave through their charged-current interactions (for a detailed description of the neutrino decoupling, see~\cite{Raffelt:2001kv, Fischer:2011cy}), has been proposed as one of the solutions for obtaining explosions.
The current state-of-the-art three-dimensional hydrodynamical supernova simulations~\cite{Kotake:2005zn, Takiwaki:2011db, Janka:2016fox, OConnor:2018tuw, Mezzacappa:2020oyq, Burrows:2020qrp}, which include neutrino heating but do not model neutrino flavor conversions, still have difficulties with excitability for large progenitors. This poses a question: is the lack of appropriate treatment of neutrino flavor evolution the missing piece? The answer is not yet found. To understand why one has to appreciate how challenging the task is itself. The extreme conditions inside the supernova core allow neutrinos not only to frequently interact with matter~\cite{Wolfenstein:1979ni, Mikheev:1986if} but also other neutrinos~\cite{1987ApJ...322..795F, Notzold:1987ik, Sigl:1993ctk} which makes it a highly non-linear many-body problem with $10^{58}$ neutrinos.  

To simplify the problem several approximations and assumptions have been invoked. One of the most crucial is including only one-body neutrino interaction by treating the problem as a probe neutrino interacting with a suitably chosen average over all the background neutrinos, i.e., the mean-field approximation~\cite{Balantekin:2006tg, Duan:2009cd, Duan:2010bg, Chakraborty:2016yeg, Tamborra:2020cul}. 
In this approximation it has been demonstrated that the inclusion of neutrino-neutrino interaction, charged-current and neutral-current collisions, and neutrino advection leads to changes in the process of neutrinos decoupling from matter, which in turn defines the efficiency of neutrino heating and can impact the fate of the star~\cite{Shalgar:2022rjj, Shalgar:2022lvv}. 

However mean-field approximation, which can be derived from the path integral representing the exact evolution of the many-neutrino system using stationary phase approximation~\cite{Balantekin:2006tg}, represents a significant truncation of the Hilbert space of the problem. In the mean-field approximation neutrinos remain unentangled even though they interact with each other via neutrino-neutrino coherent scattering. Increasingly more attention is being paid in the literature to understanding the amount of entanglement between individual neutrinos in many-neutrino systems present in astrophysical settings  
\cite{Bell:2003mg, Friedland:2003dv, Friedland:2003eh, Friedland:2006ke, cervia:2019, Roggero:2021asb, Patwardhan:2021rej, Cervia:2022pro, Lacroix:2022krq}. In order to calculate the entanglement between interacting neutrinos one needs 
to calculate the density matrix arising from the time evolution of the full many-body Hamiltonian associated with this many neutrino system. Over the last decade this problem was treated using techniques ranging from Bethe-ansatz framework, to standard Range-Kutta techniques and tensor network approach for a relatively small number of neutrinos   
\cite{Pehlivan:2011hp, Pehlivan:2014zua, Birol:2018qhx, Cervia:2019nzy, Patwardhan:2019zta, Rrapaj:2019pxz, Patwardhan:2021rej, Cervia:2022pro, Lacroix:2022krq}. Using near-term noisy quantum computers is also explored 
\cite{Hall:2021rbv, Yeter-Aydeniz:2021olz, Jha:2021itm, Illa:2022zgu, Amitrano:2022yyn}.

All the studies mentioned above has considered the many-body treatment of the neutrino flavor evolution in the two-flavor limit. Three-flavor case has been investigated but only within mean-field approximation~\cite{Fogli:2008fj, Duan:2008prl, Dasgupta:2008prd, Dasgupta:2009prl, Dasgupta:2010prd, Friedland:2010prl, Airen:2018nvp, Chakraborty:2019wxe, Shalgar:2021wlj} in which several significant differences from the two flavor approximations have been observed. The purpose of our paper is to treat the many-body picture using all three neutrino flavors. 

 Exact three-flavor treatment of collective neutrino oscillations is not only important for understanding the astrophysical scenarios better but also is crucial from the quantum information perspective. The three-level systems (three flavors of neutrinos in present work) are represented by qutrits (generalization of qubits to three-level systems) in quantum information science. Several algorithms that seem straightforward in case of qubits become rather complex while treating qutrits. For instance, the multipartite entanglement measures are a very active area of research in qubit systems~\cite{Xie:2021, Li:2022, Qian:2018} but are found to be very complicated to establish in case of qutrits. On the other hand, qutrits are expected to be more powerful units for computational purposes~\cite{Kaszlikowski:2000prl, Walborn:2006prl}. Applications of qutrits (and qudits in general) in quantum simulations of physics problems are also being explored~\cite{Gustafson:2021prd, Gustafson:2022arxiv}. It was suggested that entangled qutrits are less affected by noise than the entangled qubits~\cite{Collins:2002prl,Tsokeng:2018jpc}. It might be interesting to see how it affects the coherent forward scattering in neutrinos.

This paper is organized as follows. In Sec.~\ref{sec:Hamiltonian} we briefly describe the used formalism, and discuss the assumed approximations.
In Sec.~\ref{sec:Entanglement-and-Pvector} we describe how to quantify the entanglement in the many-body quantum system.
Our results for three and five neutrino systems in mixed and pure initial flavor states are presented in Sec.~\ref{sec:Results}. Finally, we summarize and discuss our results in the context of mean-field and two-flavor many body approximations in Sec.~\ref{sec:Conclusions}.

\section{The neutrino Hamiltonian}
\label{sec:Hamiltonian}

To describe the flavor evolution of an ensemble of $N$ neutrinos in a many-body picture, the Hamiltonian should have terms accounting for the vacuum oscillations, interaction of neutrinos with background matter (Mikheyev-Smirnov-Wolfenstein (MSW) matter effects~\cite{Wolfenstein:1977ue, Mikheev:1986if}) and the neutrino-neutrino interaction~\cite{1987ApJ...322..795F, Notzold:1987ik, Sigl:1993ctk}. Following Refs.~\cite{Pehlivan:2011hp, Pehlivan:2014zua, Birol:2018qhx,Cervia:2019nzy, Patwardhan:2019zta, Rrapaj:2019pxz, Patwardhan:2021rej, Cervia:2022pro, Lacroix:2022krq} we consider a scenario where the neutrino-neutrino interaction is dominant such that we can ignore the term corresponding to neutrino interaction with the background matter. To further simplify the calculations, we consider a system of neutrinos only (no antineutrinos). Therefore, the Hamiltonian can be written as a sum of vacuum $(H_{\nu})$ and self-interaction $(H_{\nu\nu})$ terms given as
\begin{equation}
    H=H_{\nu}+H_{\nu\nu} \ .
\end{equation}
Following Ref.~\cite{Pehlivan:2014zua} we write down the vacuum term in mass basis as
\begin{equation}\label{eq:vac}
    H_{\nu} = \sum_{\Vec{p}}\sum_{i=1}^{3}\sqrt{p^{2}+m_{i}^{2}} \; T_{ii}(p,\Vec{p}) \ ,
\end{equation}
where $p=|\Vec{p}|$ and $m_{i}$ denote the magnitude of the momentum and the mass of the neutrino, respectively. The neutrino bilinear $T_{ij}$, which is the generalization of the isospin formalism to three-flavor case, is given by
\begin{equation}\label{eq:Tij}
    T_{ij}(p,\Vec{p})=a_{i}^\dagger(\Vec{p})a_{j}(\Vec{p}) \ ,
\end{equation}
where $a_{i}^\dagger, a_{j}$ with $i,j=1,2,3$ are the neutrino creation and annihilation operators in the mass basis. $T_{ij}$ satisfies the following commutation relation
\begin{eqnarray}
    [T_{ij}(p,\Vec{p})&,&T_{kl}(p',\Vec{p'})]\nonumber\\
    &=&\delta_{p,E'}\delta_{\Vec{p},\Vec{p'}}\left(\delta_{kj}T_{il}(p,\Vec{p})-\delta_{il}T_{kj}(p,\Vec{p})\right) \ , \;
\end{eqnarray}
i.e., $T_{ij}$ span the SU(3) algebra.

In the ultrarelativistic approximation, Eq.~\eqref{eq:vac} becomes
\begin{equation}\label{eq:vac1}
    H_{\nu}=\sum_{\Vec{p}}\sum_{i=1}^{3}\sum_{j(\neq i)}\frac{\Delta m_{ij}^{2}}{2E} T_{ii}(p, \Vec{p}),
\end{equation}
where $\Delta m_{ij}^{2} = m_{i}^{2}-m_{j}^{2}$. In the following, we will also assume that $p \approx E$ due to the smallness of the active neutrino masses. 
The Hamiltonian from Eq.~\eqref{eq:vac1} can be transformed into flavor basis employing the unitary matrix called 
Pontecorvo–Maki–Nakagawa–Sakata (PMNS) mixing matrix~\cite{Pontecorvo:1957qd, Maki:1962mu} parameterized by three mixing angles $\theta_{12},~\theta_{13},~\theta_{23}$ and a single $\delta_\mathrm{CP}$ CP-violating Dirac phase. In the absence of sterile neutrinos this phase can be factorized out of the total Hamiltonian~\cite{Pehlivan:2014zua} and does not affect the non-linear flavor transformations streaming from neutrino-neutrino interactions, therefore we ignore $\delta_\mathrm{CP}$ in our calculations. The values of the other five  parameters are given in Table~\ref{tab:param}~\cite{ParticleDataGroup:2022pth}.

The collective oscillation term which leads to nonlinear dynamics in flavor evolution, can be written as~\cite{Pehlivan:2014zua}
\begin{eqnarray}
    H_{\nu\nu}&=&\frac{G_{F}}{\sqrt{2}V}\sum_{i,j=1}^{3}\sum_{E,\Vec{p}}\sum_{E',\Vec{p'}}\left(1-\cos\theta_{\Vec{p}\Vec{p'}}\right)\\
    &~&\times T_{ij}(E,\Vec{p})T_{ji}(E',\Vec{p'}) \ ,
\end{eqnarray}
where $G_{F}$ is the Fermi constant, $V$ is the quantization volume, and $\theta_{\Vec{p}\Vec{p'}}$ is the angle between neutrinos with momenta $\Vec{p}$ and $\Vec{p'}$. $H_{\nu\nu}$ is rotationally invariant and hence the same in both mass and flavor basis~\cite{Pehlivan:2014zua}.

To further simplify the form of Hamiltonian, we write $T_{ij}$ in terms of SU(3) generators. Hence, Eq.~\eqref{eq:Tij} can be written in the following form (we drop $E$ and $p$ for readability)
\begin{eqnarray}
    T_{ij}=\sum_{i'}(\lambda_{i'})_{ji}Q_{i'}+\frac{1}{3}\delta_{ij}\sum_{i}a_{i}^\dagger a_{i},
\end{eqnarray}
where $\lambda$'s are the Gell-Mann matrices, and the generators $Q_{i'}$ are given by
\begin{equation}
    Q_{i'}=\frac{1}{2}\sum_{i,j=1}^{3}a^{\dagger}_{i}(\lambda_{i'})_{ij}a_{j} \ ,
\end{equation}
for instance, $Q_{8 } =\frac{1}{2\sqrt{3}}(a_{1}^{\dagger}a_{1}+a_{2}^{\dagger}a_{2}-2a_{3}^{\dagger}a_{3})$. 

Hence, we can write the total Hamiltonian in a compact form:
\begin{equation}\label{eq:H}
    H = \sum_{p}\Vec{B}\cdot\Vec{Q_{p}}+\sum_{p,p'}\mu_{pp'}\Vec{Q_{p}}\cdot\Vec{Q_{p'}} \ ,
\end{equation}
where the neutrino-neutrino interaction strength parameter is 
\begin{equation}
\label{eq:interaction-strenght}
    \mu_{pp'} = \frac{\sqrt{2}G_{F}}{V}(1-\cos\theta_{pp'}) \ ,
\end{equation}
and the used auxiliary vector is given by
\begin{equation}
    \Vec{B}=\left(0,0,\omega_{p},0,0,0,0,\Omega_{p}\right) \ .
\end{equation}
Here the oscillation frequencies are
\begin{subequations}
\label{eq:constants} 
\begin{eqnarray}
    \omega_{p}&=&-\frac{1}{2E} \delta m^{2} \ , \label{eq:wp} \\    
    \Omega_{p}&=&-\frac{1}{2E}\Delta m^{2} \label{eq:Wp} \ ,
\end{eqnarray}
\end{subequations}
where $\delta m^{2} = m_{2}^{2} - m_{1}^{2}$, and $\Delta m^{2} \approx |m_{3}^{2}-m_{2}^{2}| \approx |m_{3}^{2}-m_{1}^{2}|$. 
The sign of the smaller mass squared difference $\delta m^{2}$ has been determined by the observation of the MSW matter effect in the solar neutrino data~\cite{SNO:2001kpb}. The sign of bigger mass squared difference $\Delta m^{2}$ -- mass ordering --- is, however, still unknown. In the next decade, the existing and upcoming experiments that are looking at atmospheric and accelerator neutrinos such as DUNE~\cite{ DUNE:2020ypp}, JUNO~\cite{JUNO:2015zny, IceCube-Gen2:2019fet}, Hyper-Kamiokande~\cite{Hyper-Kamiokande:2018ofw}, Ice-Cube~\cite{IceCube-Gen2:2019fet}, and KM3NeT~\cite{KM3NeT:2021ozk} are expected to determine the mass ordering with 3$\sigma$ - 5$\sigma$ significance. The current results from cosmological observations seem to favor the normal mass ordering based on the measurements of the sum of the neutrino masses $\sum_i m_{\nu,i} \lesssim 0.1~\mathrm{eV}$~\cite{Zyla:2020zbs}. However, these results are dependent on the priors used in the statistical analyses~\cite{Gariazzo:2018pei, Jimenez:2022dkn, Gariazzo:2022ahe}. Therefore, in Sec.~\ref{sec:Results}, we present the results for both mass orderings.

The geometric term, which depends on the angle between the trajectories of the two neutrinos $\theta_{pp'}$, present in the neutrino-neutrino interaction strength parameter $\mu_{pp'}$ leads to the complexities in the collective neutrino oscillations problem even in the simplest approximations like mean-field calculations~\cite{Balantekin:2006tg, Duan:2009cd, Duan:2010bg, Chakraborty:2016yeg, Tamborra:2020cul}.
To make this problem tractable, the so-called ``single-angle" approximation has been adopted heavily in literature (see, e.g.~Refs.~\cite{Qian:1994wh, Bell:2003mg, Friedland:2006ke}).
In ``single-angle" approximation, an interaction strength averaged over all the possible angles is considered. For instance, in the ``neutrino bulb" model~\cite{Duan:2006an}, which we also adopt, the neutrino-neutrino interaction strength takes the following form
\begin{equation}
\label{eq:mu-nunu}
    \mu(r) = \frac{G_{F}}{\sqrt{2}V}\left(1-\sqrt{1-\frac{R_{\nu}^{2}}{r^{2}}}\right)^{2} \ ,
\end{equation}
where $r$ is the distance from the center of the star, and $R_{\nu}$ is the radius of the neutrinosphere; a radius at which the probability of a single neutrino interaction with the other matter particles present in the medium drops below unity. 
 
 In order to make our approach and calculations more readily comparable with the existing works on the many-body effects in the two flavor case~\cite{Pehlivan:2011hp, Pehlivan:2014zua, Birol:2018qhx, Cervia:2019nzy, Patwardhan:2019zta, Rrapaj:2019pxz, Patwardhan:2021rej, Cervia:2022pro, Lacroix:2022krq}, we use a similar convention, i.e., we consider the $\omega_{p}$ and $\Omega_{p}$ in units of $\kappa$, a suitably chosen scaling factor.
 We consider a system of $N$ neutrinos with discrete momenta that correspond to frequencies $\omega_{q} = q \omega_{p} $ and $\Omega_{q} = q \Omega_{p} $; we denote the frequency mode by $q=1,2,\ldots, N$, and  $\omega_{p}$ and $\Omega_{p}$ are given by Eq.~\ref{eq:constants}. 
All parameters used in the calculations are given in \tableautorefname~\ref{tab:param}.

\begin{table}[h!]
    \centering
    \caption{Parameters used in the calculations.}
    \label{tab:param}
    
    \begin{tabular}{|c|c|c|}
    \hline
       Parameter  &  Value\\\hline
        $\kappa$ & $10^{-17}$ MeV\\
        $\delta m^{2}$ & $7.59\times 10^{-17}$ MeV$^{2}$\\
        $\Delta m^{2}$ & $\pm2.32\times 10^{-15}$ MeV$^{2}$\\
        $E$ & $10$ MeV\\
        $R_{\nu}$ & $32.2\kappa^{-1}$\\
        $\mu(R_{\nu})$ & $3.62\times10^{4} \kappa$\\
        $\theta_{12}$ & $33.44^{\circ}$ \\
        $\theta_{13}$ & $8.57^{\circ}$ \\
        $\theta_{23}$ & $49.2^{\circ}$ \\\hline
    \end{tabular}
\end{table}
\section{Quantum entanglement in many-body systems}
\label{sec:Entanglement-and-Pvector}

In this section we define the equation of motion in terms of the polarization vectors (Sec.~\ref{sec:Polarization-vector}) and introduce the concept of the entanglement between neutrinos (Sec.~\ref{sec:Entangelement}).

\subsection{Density matrix and Polarization vector formalism}
\label{sec:Polarization-vector}

The density matrix of a pure quantum system can be written  as follows
\begin{equation}
\label{eq:density-matrix}
    \rho \equiv |\Psi \rangle \langle \Psi | \ ,
\end{equation}
where the outer product of the state vector ($|\Psi \rangle$) of the system with itself is taken. 
If we ignore the interactions of neutrinos with other particles, the state $\ket{\Psi}$ represents the entire $N$ neutrino system. 
The reduced density matrix for a single neutrino (labeled $n$) in a system with three flavors can be written in terms of the fundamental elements of the SU(3) algebra representation -- $\lambda$ matrices -- as follows
\begin{equation}
\label{eq:reduced-density-matrix}
\rho_n = \mathrm{Tr}_{n}[\rho] = \frac{1}{3} \left[\mathbb{I} + \frac{3}{2}\sum_j\lambda_j P_j \right] \ ,
\end{equation}
where $\mathbb{I}$ is the $3\times 3$ identity matrix, and $P_j$ is the $j$-th component of the polarization vector $\vec{P}$ of the dimension $8\times1$.
For the three level systems, such as ours, the $\lambda$ matrices are usually taken to be the $3\times3$ Gell-Mann matrices. 

We can extract the polarization vector components in the three-flavor scenario from the density matrix as 
\begin{equation}
\label{eq:Pi}
P_{j} = {\rm Tr}[\rho_n \lambda_{j}] \ .
\end{equation}
In Eq.~\eqref{eq:Pi}, we have employed the following property of Gell-Mann matrices
\begin{equation}
\label{eq:identity}
    \lambda_{i}\lambda_{j} = \frac{2}{3}\delta_{ij}\mathbb{I} + (d_{ijk}+if_{ijk})\lambda_{k} \ .
\end{equation}
In the three-flavor case, the two components of the total polarization vector (flavor-lepton numbers) $P_{3}$ and $P_{8}$ are conserved separately. Whereas, in the two-flavor case only one component of the total polarization vector $(P_{3})$ is conserved. The $P_{3}$ and $P_{8}$ values for each neutrino vary from $-1$ to $1$ and from $-2/\sqrt{3}$ to $1/\sqrt{3}$, respectively. Similar to two-flavor case, extrema of the components of the individual polarization vector values represent the pure mass eigenstates represented by $\nu_{1},\nu_{2}$ and $\nu_{3}$ (see \tableautorefname~\ref{tab:p38}) without any entanglement. 
In the two-flavor scenario (two level system) all the absolute values of the polarization vector are allowed because the Lie algebras of SU(2) and SO(3) are isomorphic; all points on the Bloch sphere are viable solutions for a pure state. In the three-flavor, however, that is no longer true because SU(3) and SO(8) are not isomorphic, i.e., only certain solutions are allowed (see also Appendix~\ref{app:entropy}).\\

\begin{table}[h!]\caption{The $P_{3}$ and $P_{8}$ values corresponding to the three pure mass eigenstates.}
    \centering
    \begin{tabular}{|c|c|c|}
    \hline
       Mass eigenstate &  $P_{3}$ &  $P_{8}$\\\hline
        $\nu_{1}$ & $1$ & $1/\sqrt{3}$\\
        $\nu_{2}$ & $-1$ & $1/\sqrt{3}$\\
        $\nu_{3}$ & $0$ & $-2/\sqrt{3}$\\\hline
    \end{tabular}
    \label{tab:p38}
\end{table}

We can also write the probability of finding a neutrino in mass eigenstates $\nu_{1},\nu_{2}$ and $\nu_{3}$ in terms of $P_{3}$ and $P_{8}$ values as follows:
\begin{subequations}
\label{eq:P_nui} 
\begin{eqnarray} \label{eq:prob}
P_{\nu_{1}}&=&\frac{1}{3}\left(1+\frac{3}{2}P_{3}+\frac{\sqrt{3}}{2}P_{8}\right)\ , \\
P_{\nu_{2}}&=&\frac{1}{3}\left(1-\frac{3}{2}P_{3}+\frac{\sqrt{3}}{2}P_{8}\right)\ , \\
P_{\nu_{3}}&=&\frac{1}{3}\left(1-\sqrt{3}P_{8}\right)\ .
\end{eqnarray}
\end{subequations}
We will utilize these probabilities to understand the mixing of different mass eigenstates as the system evolves.

\subsection{Entanglement measures}
\label{sec:Entangelement}

To quantify the entanglement, we divide the system in two parts, $n$th neutrino as one part and the remaining neutrinos as the second part. Therefore, the total Hilbert space for $N$ neutrino system can be written in terms of $n$th neutrino Hilbert space $\mathcal{H}_{n}$ and the remaining one as $\mathcal{H}_{N}=\mathcal{H}_{n}\otimes\mathcal{H}_{N-n}$.
In such a system we can calculate the bipartite entanglement entropy for each neutrino with 
\begin{equation}
\label{eq:bipartite-entropy}    
S_{n} = - \mathrm{Tr} \left[ \rho_n \log{\rho_n} \right] \ ,
\end{equation}
where $\rho_{n}$ is the reduced density matrix calculated as
\begin{eqnarray}
\rho_{n}={\rm Tr}_{N-n}[\rho].
\end{eqnarray}
Equivalently, Eq.~\eqref{eq:bipartite-entropy} can also be written in terms of eigenvalues $(\beta_i)$ of the reduced density matrices $\rho_{n}$
\begin{equation}
\label{eq:bipartite-entropy_eigenvalues}    
S_n = - \sum_{i}\beta_{i} \log{\beta_{i}} \ .
\end{equation}
For three-flavor case, using the reduced density matrix given in Eq.~\eqref{eq:reduced-density-matrix}, the entanglement entropy can be also expressed as
\begin{equation}
\label{eq:entropy-Baha}
    S = \log{3} - \frac{1}{3} \mathrm{Tr} \left[\left(\mathbb{I} + \frac{3}{2}\lambda_j P_j\right) \log\left(\mathbb{I} + \frac{3}{2}\lambda_j P_j \right) \right] \ .
\end{equation}
The exact solution for entropy in terms of the two SU(3) invariants, namely the magnitude of the eight-dimensional polarization vector $|\vec{P}|$ and  $\Pi=d_{ijk}P_{i}P_{j}P_{k}$ is given in \appendixautorefname~\ref{app:entropy}.

We can show that the entanglement entropy expressed in Eq.~\eqref{eq:entropy-Baha} is always positive and bounded from above.
Using the inequality
\begin{equation}
    \frac{x}{1+x} <  \log(1+x)< x, 
\end{equation}
valid for $x>-1$ and $x\neq 0$, we can write for each eigenvalue $x_{a}$ of $(3/2)\lambda_{j}P_{j}$,
\begin{equation}
    x_{a}<(1+x_{a})\log(1+x_{a}). 
\end{equation}
Adding these equations for all the eigenvalues and noting that the trace of $\lambda_{j}P_{j}$ is $0$, we obtain
\begin{equation}
    \mathrm{Tr}\left[\left( \mathbb{I}  + \frac{3}{2}\lambda_{j}P_{j}\right)\log\left(\mathbb{I} + \frac{3}{2}\lambda_{j}P_{j} \right)\right] > 0 \ .
\end{equation}
It is straightforward to notice that for a maximally entangled state the entropy is equal to $\log 3$ in three-flavor case ($\log 2$ in two-flavor case).
In addition, as can be seen from Eq.~\eqref{eq:entropy-Baha}, the entropy and the polarization vector components have a direct correlation. Therefore, similarly as in the case of the entropy, we can examine the entanglement by looking at the polarization vector. In the two flavor case, only a single component of the polarization vector, $P_{3}$ carries the information about the entanglement. However, in the present case of three flavor, there are two components $P_{3}$ and $P_{8}$ because two of the Gell-Mann matrices are diagonal, which makes the analysis of the entanglement more challenging as compared to the two flavor case.

\section{Results}
\label{sec:Results}
\begin{figure}[t]
    \centering
    \includegraphics[width=0.99\linewidth]{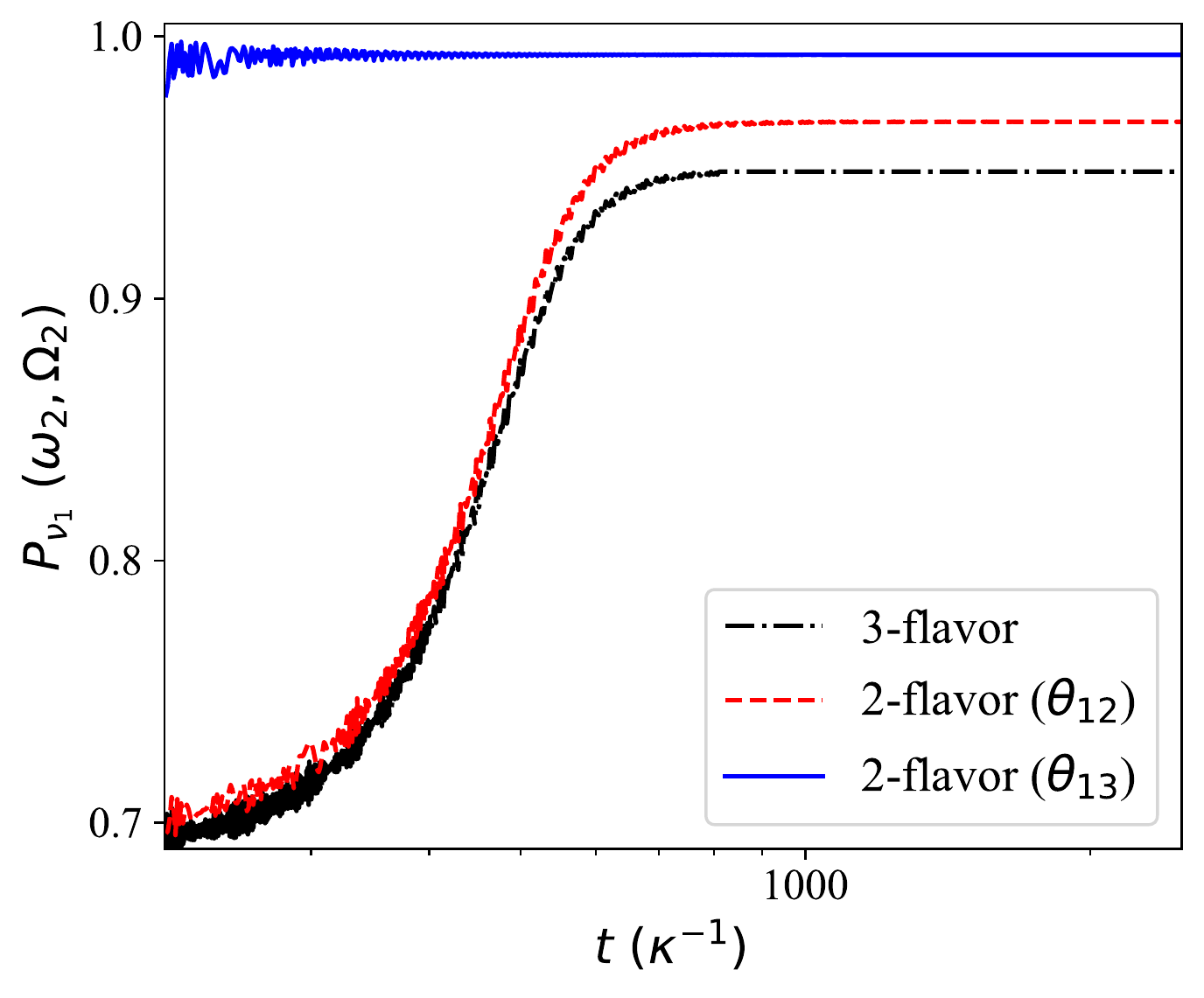}
    \caption{The comparison of the probability of finding the $q=2$ frequency mode neutrino in $\nu_{1}$ mass eigenstate ($P_{\nu_{1}}$) in 3-flavor and 2-flavor case. Different mixing angles $\theta_{12}$ and $\theta_{13}$ are considered in 2-flavor case. The results are for $N=5$ neutrinos with initial state $\ket{\psi}=\ket{\nu_{e}\nu_{e}\nu_{e}\nu_{e}\nu_{e}}$ in NO.}
    \label{fig:N5_comp}
\end{figure}

We are interested in the time evolution of neutrino flavor under the Hamiltonian given in \sectionautorefname~\ref{sec:Hamiltonian}. We solve the time-dependent Schr\"odinger equation numerically using fourth order Runge-Kutta (RK4) method. As can be seen from Eqs.~\eqref{eq:H} and \eqref{eq:mu-nunu}, the Hamiltonian in our case is time dependent, and hence the numerical complexities are enhanced making our calculations limited to a small number of neutrinos. We solve our equations in flavor basis by transforming the Hamiltonian given in mass basis (Eq.~\eqref{eq:H}) by utilizing the unitary PMNS mixing matrix. Then, at the end, we transform the evolved wave function to mass basis and calculate the polarization vector components $P_{3}$ and $P_{8}$ in this basis for a better presentation.

\begin{figure*}[t]
    \centering
    \includegraphics[width=0.975\columnwidth]{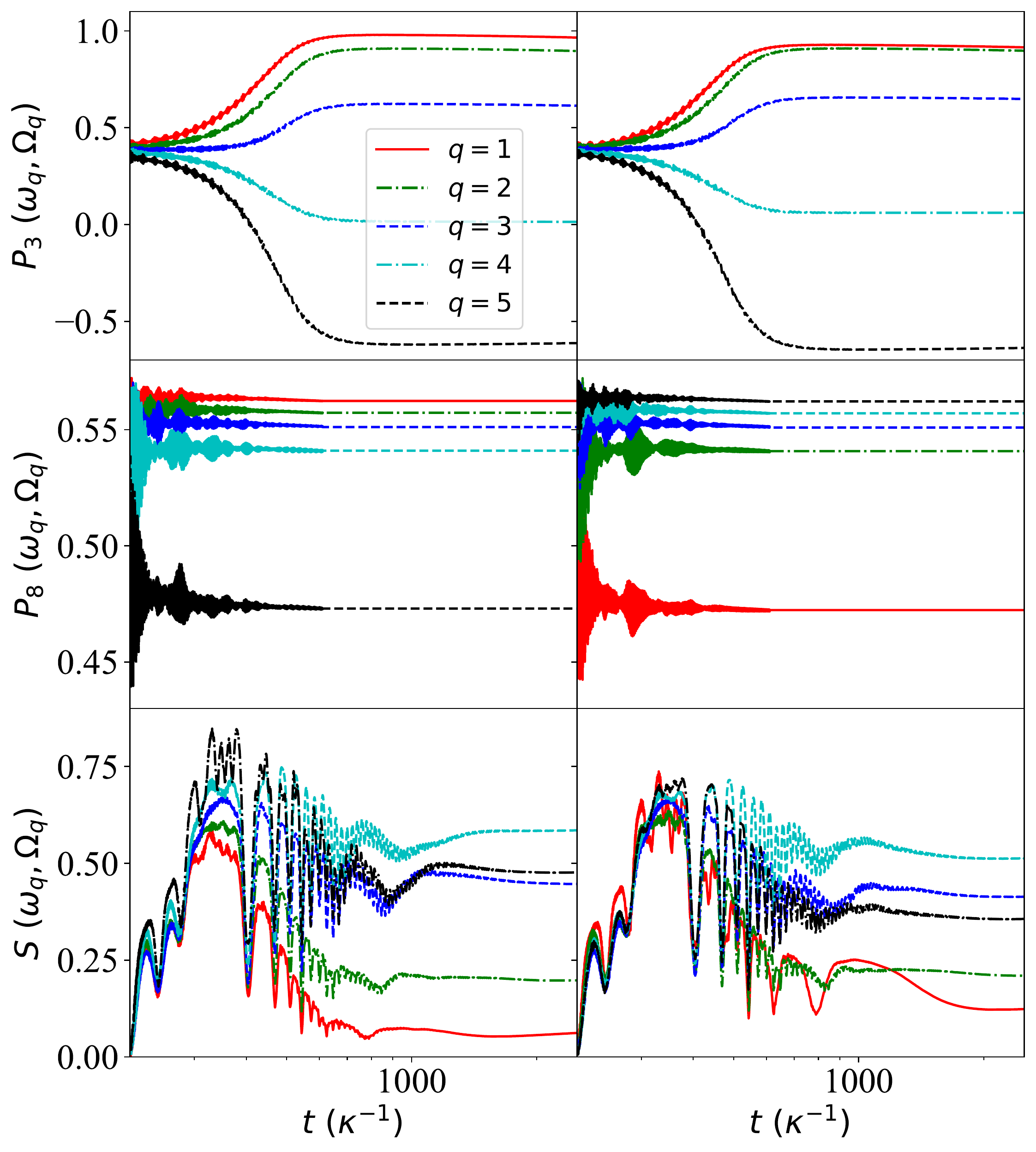}
    \includegraphics[width=0.99\columnwidth]{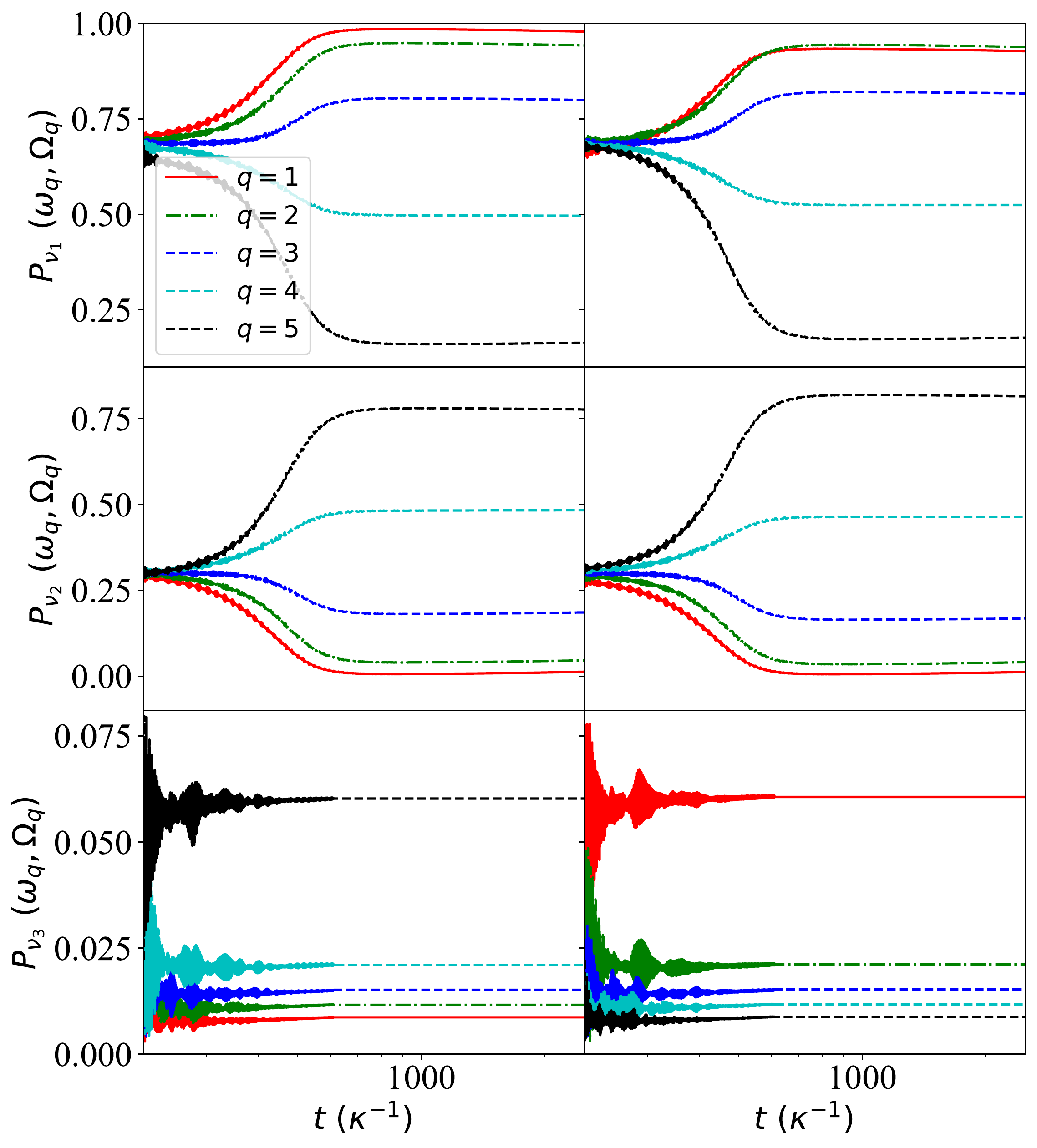}

    \caption{{\it Left panels:} The temporal evolution of $P_{3}$ (top), $P_{8}$ (middle) and $S$ (bottom) for $N=5$ neutrinos with initial state $\ket{\psi}=\ket{\nu_{e}\nu_{e}\nu_{e}\nu_{e}\nu_{e}}$ in NO [left] and IO [right]. {\it Rights panels:} The temporal evolution of $P_{\nu_{1}}$ (top), $P_{\nu_{2}}$ (middle) and $P_{\nu_{3}}$ (bottom) for $N=5$ neutrinos with initial state $\ket{\psi}=\ket{\nu_{e}\nu_{e}\nu_{e}\nu_{e}\nu_{e}}$ in NO [left] and IO [right].}
    \label{fig:N5}
\end{figure*}

To demonstrate the importance of considering three-flavor case instead of the two-flavor approximation, we compute the probability of a neutrino to be found in one of the mass eigenstates. For a comparison, we consider a simple example of a five neutrinos system with an initial state having all electron neutrinos $\ket{\nu_{e}\nu_{e}\nu_{e}\nu_{e}\nu_{e}}$. The $P_{\nu_{1}}$ for the neutrino in frequency mode $q=2$ are shown in \figureautorefname~\ref{fig:N5_comp}. In the asymptotic limits, the $P_{\nu_{1}}$ probabilities in two-flavor case are always larger as compared to the three-flavor case. It signifies more flavor mixing in three flavor case and hence more entanglement, irrespective of the value of the mixing angle considered in two-flavor approximation. These differences in mixing and entanglement further enhance in case of an initial state with mixed flavors (see Sec.~\ref{sec:mixed}). Therefore, a proper treatment of the three flavor neutrino evolution is worth further exploring.

In this section, we examine the evolution of $P_{3}$ and $P_{8}$ components of the polarization vector and the entanglement entropy in systems made of three and five neutrinos to investigate the underlying connections. The probabilities of finding a neutrino in a particular mass eigenstate are also investigated to understand the mixing of different mass eigenstates. First, in Sec.~\ref{sec:pure} we look at the system of five neutrinos all with pure electron flavor in the initial state, then in Sec.~\ref{sec:mixed} we investigate the evolution of systems with three and five neutrinos in mixed initial states. In Sec.~\ref{sec:triangle-plots} we explore the behavior of entanglement in terms of $P_{3}$ and $P_{8}$ components in $\hat{e}_{3}$-$\hat{e}_{8}$ plane. We perform all of our calculations for both normal (NO) and inverted (IO) neutrino mass orderings.

\subsection{The evolution of a neutrino system with all-electron flavor initial state}
\label{sec:pure}

\begin{figure*}[t]
    \centering
    \includegraphics[width=0.49\linewidth]{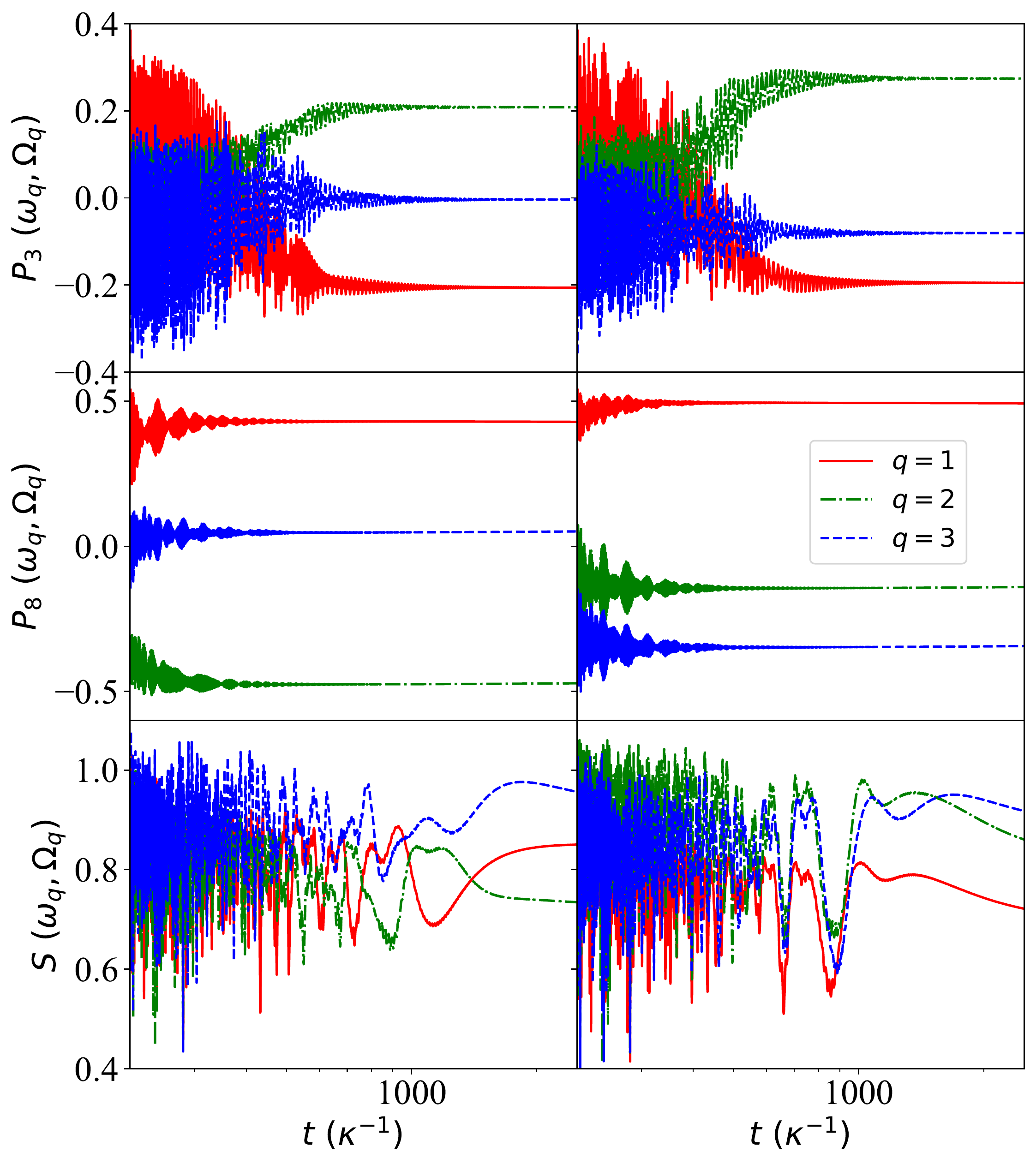}
    \includegraphics[width=0.4805\linewidth]{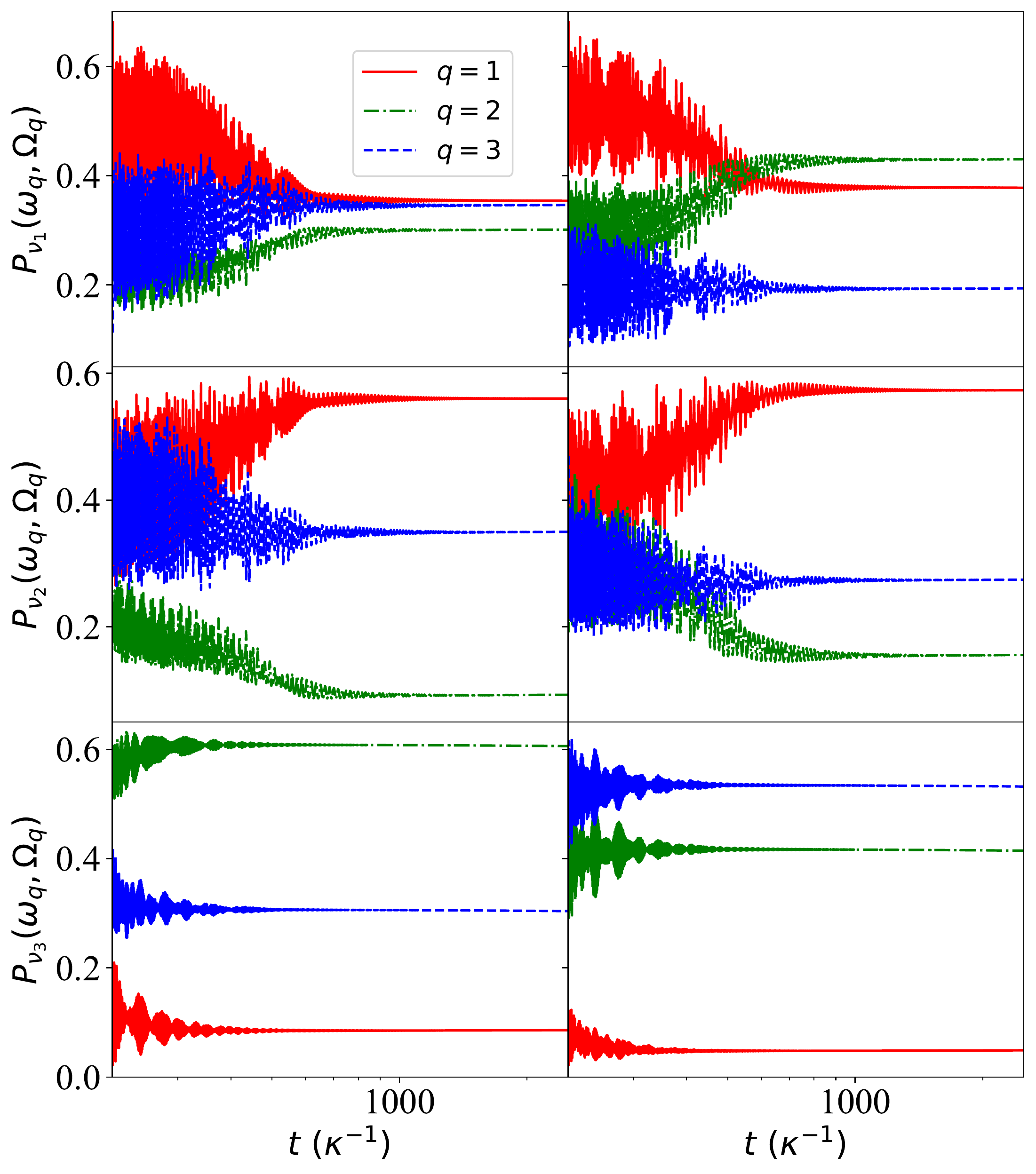}
    \caption{{\it Left panels:} The temporal evolution of $P_{3}$ (top), $P_{8}$ (middle) and $S$ (bottom) for $N=3$ neutrinos with initial state $\ket{\psi}=\ket{\nu_{e}\nu_{\mu}\nu_{\tau}}$ in NO (left) and IO (right). {\it Right panels:} The temporal evolution of $P_{\nu_{1}}$ (top), $P_{\nu_{2}}$ (middle) and $P_{\nu_{3}}$ (bottom) for $N=3$ neutrinos with initial state $\ket{\psi}=\ket{\nu_{e}\nu_{\mu}\nu_{\tau}}$ in NO (left) and IO (right). Red solid, green dashed-dotted and blue dashed lines represent first, second and third neutrino, respectively.}
    \label{fig:N3}
\end{figure*}

For an initial state with all neutrinos in electron flavor $(\nu_{e})$, the results for time evolution of five neutrino system are shown in left panels of \figureautorefname~\ref{fig:N5}. The $P_{3}$ values follow the same hierarchy in both the NO and IO. However, the hierarchy of $P_{8}$ values is reversed; the $P_{8}$ for the maximum frequency neutrino is the lowest in NO whereas highest in the case of IO. The entropies $(S)$ do not follow any particular order for this initial state in contrast to the two flavor case where the entropy of maximum frequency neutrino is maximum and so on~\cite{cervia:2019}.

In both NO and IO, the entropies at the asymptotic limits follow the same order, except for neutrinos with frequency modes $q=3$ and $q=5$, but with slightly different magnitudes. Furthermore, the entropies are significantly higher as compared to two-flavor case~\cite{cervia:2019, Cervia:2022pro} which indicates larger deviations from the mean-field approximation where the entropies are zero by default.

We can further understand the dependence of flavor mixing in different mass orderings from the probabilities of a neutrino to be found in a particular mass eigenstate shown in \figureautorefname~\ref{fig:N5} [right panels]. In the case of NO, the neutrino with lowest frequency mode $(q=1)$ is predominantly in first mass eigenstate with negligible mixing of second and third ones. The neutrino with highest frequency mode $(q=5)$ is predominantly in the second mass eigenstate and has the largest mixing of third mass eigenstate as compared to other neutrinos. In the IO, the neutrino with lowest frequency mode $(q=1)$ has the largest mixing of third mass eigenstate and predominantly in first mass eigenstate. Whereas, the highest frequency neutrino $(q=5)$ has the least contribution of third mass eigenstate in contrast to NO.

\subsection{The evolution of a neutrino system with mixed-flavor initial states}
\label{sec:mixed}

In the case of an initial state with mixed flavors, we first present the results for a system of three neutrinos in \figureautorefname~\ref{fig:N3}. We consider a state with each neutrino in different flavor, i.e., $\ket{\nu_{e}\nu_{\mu}\nu_{\tau}}$. As can be seen from the \figureautorefname~\ref{fig:N3}, the ordering of $P_{3}$ is the same in both NO and IO, but the magnitudes of $P_{3}$ corresponding to neutrino with frequency modes $q=2$ and $q=3$ are shifted towards larger absolute values in the IO. The $P_{8}$ values corresponding to neutrino with frequency mode $q=2$ and $q=3$ are flipped in the IO. The difference between the $P_{8}$ values corresponding to $q=2$ and $q=3$ is decreased by a factor of $\sim 2$. The $P_{8}$ value corresponding to $q=1$ has also increased slightly in the case of IO. We see a drastic change in the hierarchy of entropies in different mass orderings. One interesting point to note here is that the entropy corresponding to any neutrino does not reach to the maximum, {\it i.e.}, $\log(3)$. On the other hand, in two flavor approximation, the entropy reaches to the maximum value, {\it i.e.}, $\log(2)$ even for a small number of neutrinos $(N=2)$~\cite{cervia:2019}.

From the probabilities $P_{\nu_{i}}$s shown in \figureautorefname~\ref{fig:N3} [right panels], we can see that the neutrino in the lowest frequency mode $(q=1)$ has almost same probabilities for $i=1,2$ and $3$ in both mass orderings. The neutrino with $q=2$ and $3$ show drastic changes in different mass orderings. In the NO, the $q=2$ neutrino is predominantly in third mass eigenstate with the probability $\sim60\%$ and a significant mixing of first mass eigenstate. The probability for being in second mass eigenstate is very small $(\sim5\%)$. In IO, this neutrino has almost equal mixing of first and third mass eigenstates with nearly $\sim15\%$ mixing of second mass eigenstate. Similarly, the $q=3$ neutrino has almost equal contribution of all three mass eigenstates in NO. In IO, this neutrino is predominantly in third mass eigenstate and the contribution of first and second states is reduced to nearly $20\%$ and $25\%$, respectively.

\begin{figure*}[t]
    \centering
    \includegraphics[width=0.99\columnwidth]{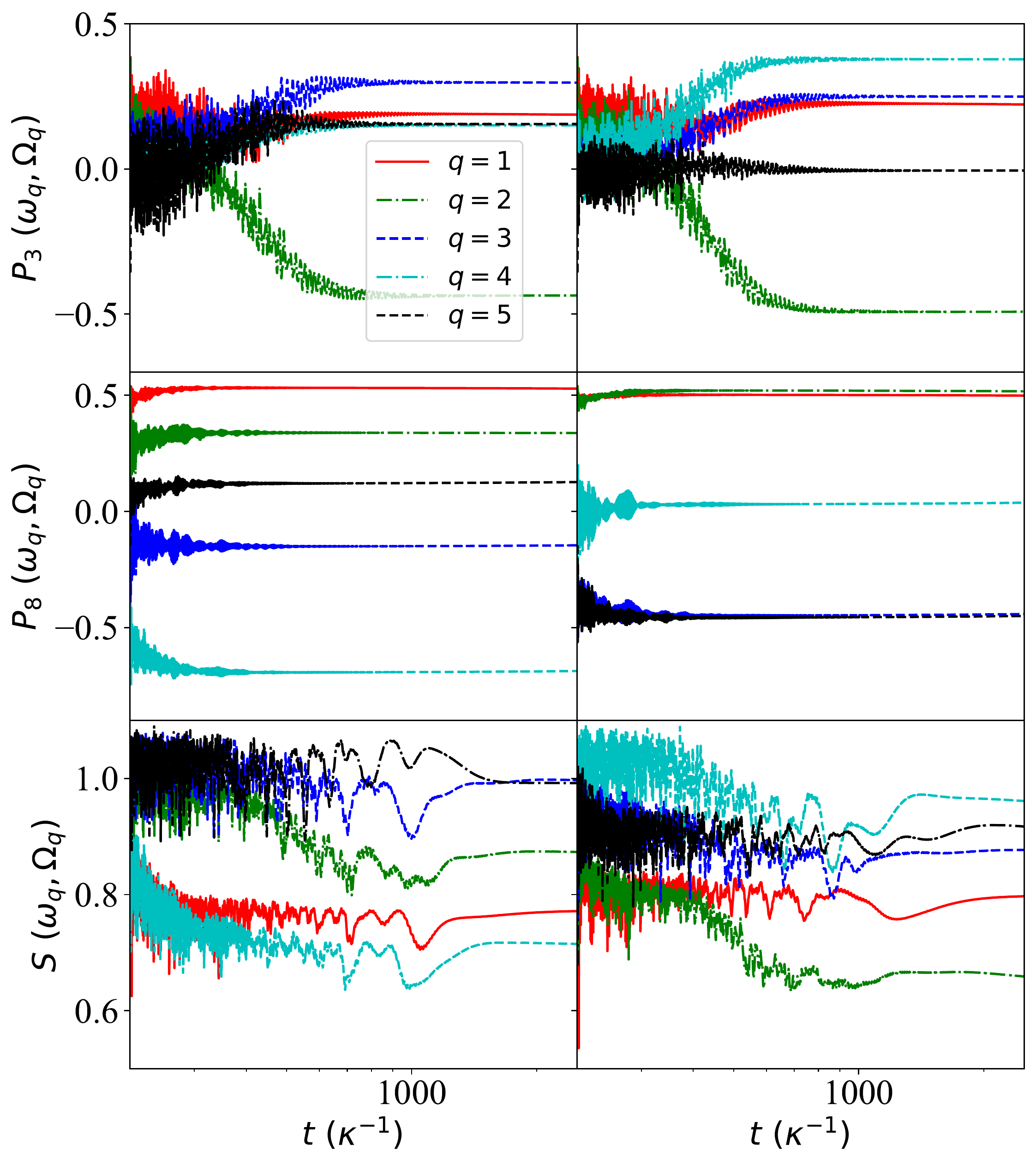}
    \includegraphics[width=0.99\columnwidth]{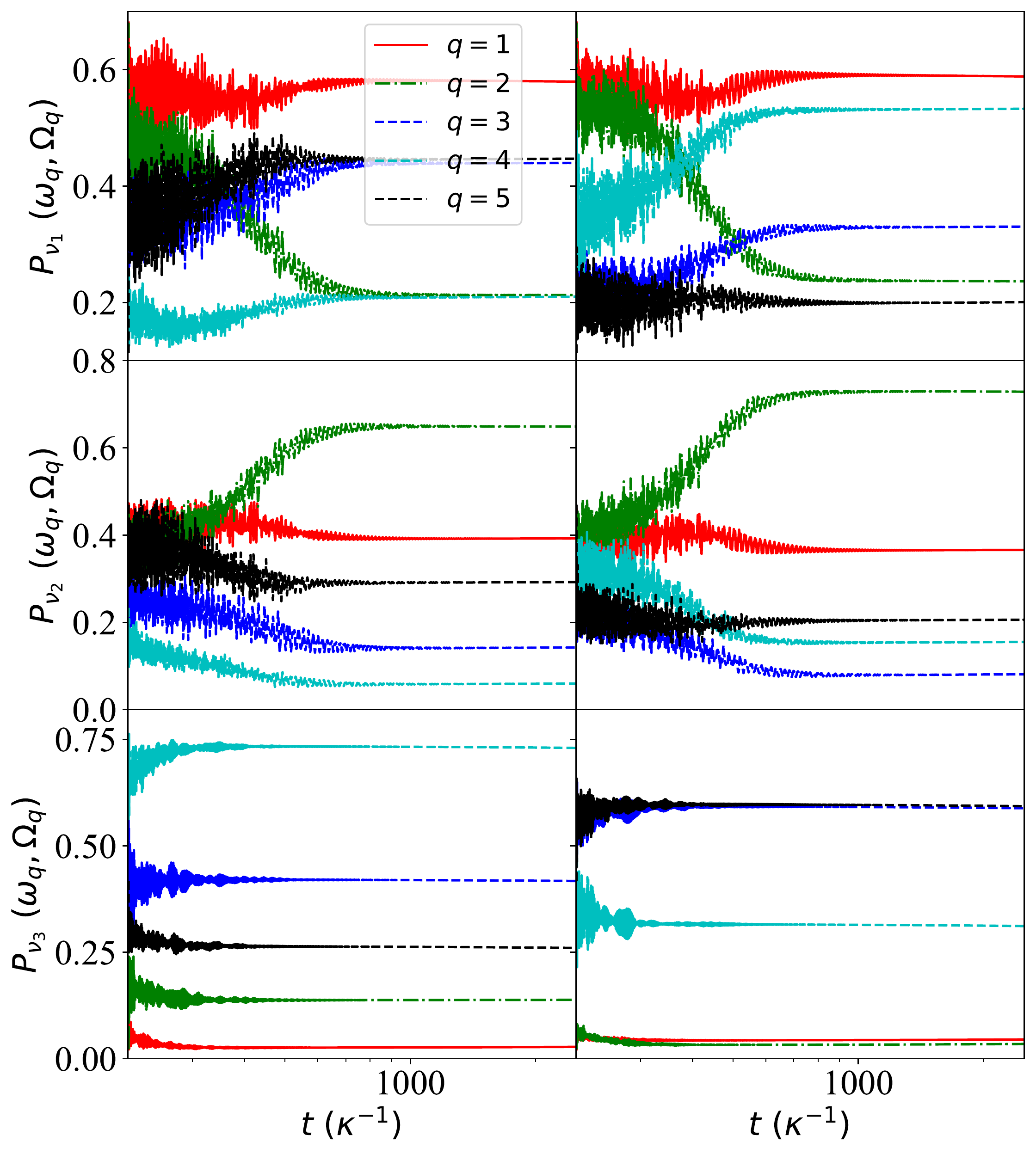}
    \caption{The temporal evolution of $P_{3}$ (top), $P_{8}$ (middle) and $S$ (bottom) for $N=5$ neutrinos with initial state $\ket{\psi}=\ket{\nu_{e}\nu_{e}\nu_{\mu}\nu_{\mu}\nu_{\tau}}$ in NO (left) and IO (right). {\it Right panels:} The temporal evolution $P_{\nu_{1}}$ (top), $P_{\nu_{2}}$ (middle) and $P_{\nu_{3}}$ (bottom) for $N=5$ neutrinos with initial state $\ket{\psi}=\ket{\nu_{e}\nu_{e}\nu_{\mu}\nu_{\mu}\nu_{\tau}}$ in NO (left) and IO (right).}
    \label{fig:N5_mix}
\end{figure*}

The results for a system of five neutrinos with an initial state of mixed flavors $\ket{\nu_{e}\nu_{e}\nu_{\mu}\nu_{\mu}\nu_{\tau}}$ are shown in \figureautorefname~\ref{fig:N5_mix}. A more prominent difference can be seen in the ordering of $P_{3}$ and $P_{8}$ values as compared to the initial state with all neutrinos in electron flavor (see \figureautorefname~\ref{fig:N5}). The entropy for the maximally entangled neutrino (the neutrino in the frequency mode $q=5$ in NO, and in frequency mode $q=4$ in IO) is slightly larger approaching to the maximum possible value {\it i.e.}, $\log(3)$ at initial times, as compared to the case of three neutrinos (see~\figureautorefname~\ref{fig:N3}). Therefore, we see that the entropy  increases further with the number of neutrinos, in agreement with what was pointed out in Ref.~\cite{cervia:2019}.

In NO, we can see from the entropies $S$ that the neutrino with $q=4$ is the least entangled one, whereas in the IO, it is the most entangled one. The neutrino in mode $q=1$ has almost similar values for $P_{3},P_{8}$ and $S$ in both mass orderings. Hence, we note that for a three flavor neutrino system the evolution becomes more complex due to mixed flavor initial states. We can get better insights from the probabilities $P_{\nu_{i}}$ shown in \figureautorefname~\ref{fig:N5_mix} [right panels]. For the neutrino in mode $q=1$, similar to the $P_{3},P_{8}$ and $S$ values, these probabilities are almost similar in both mass orderings. The neutrino in mode $q=2$ has more contribution of third mass eigenstate in NO as compared to IO. In NO, the neutrino in mode $q=4$ is predominantly in the third mass eigenstate with negligible contribution of second state. In IO, this neutrino is predominantly in the first mass eigenstate with a significant contribution of third state and non-negligible contribution of second one. Hence, the time evolution of neutrino system in three flavor has a strong dependence on the mass orderings especially in case of mixed initial state.

\subsection{Asymptotic values of the polarization vector's components in $\hat{e}_{3}$-$\hat{e}_{8}$ plane}
\label{sec:triangle-plots}

\begin{figure*}[t]
    \centering
    \includegraphics[width=0.99\columnwidth]{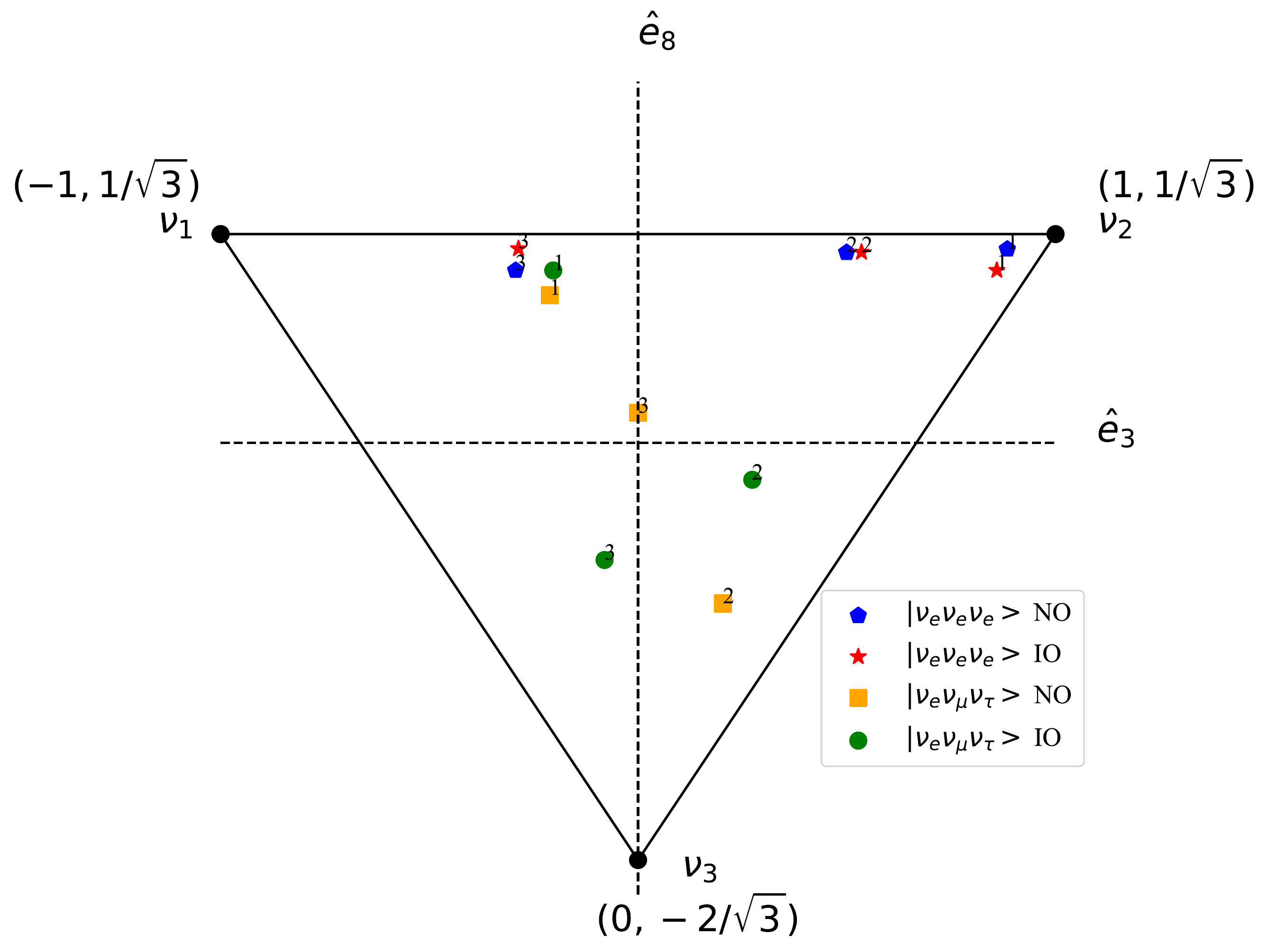}
    \includegraphics[width=0.99\columnwidth]{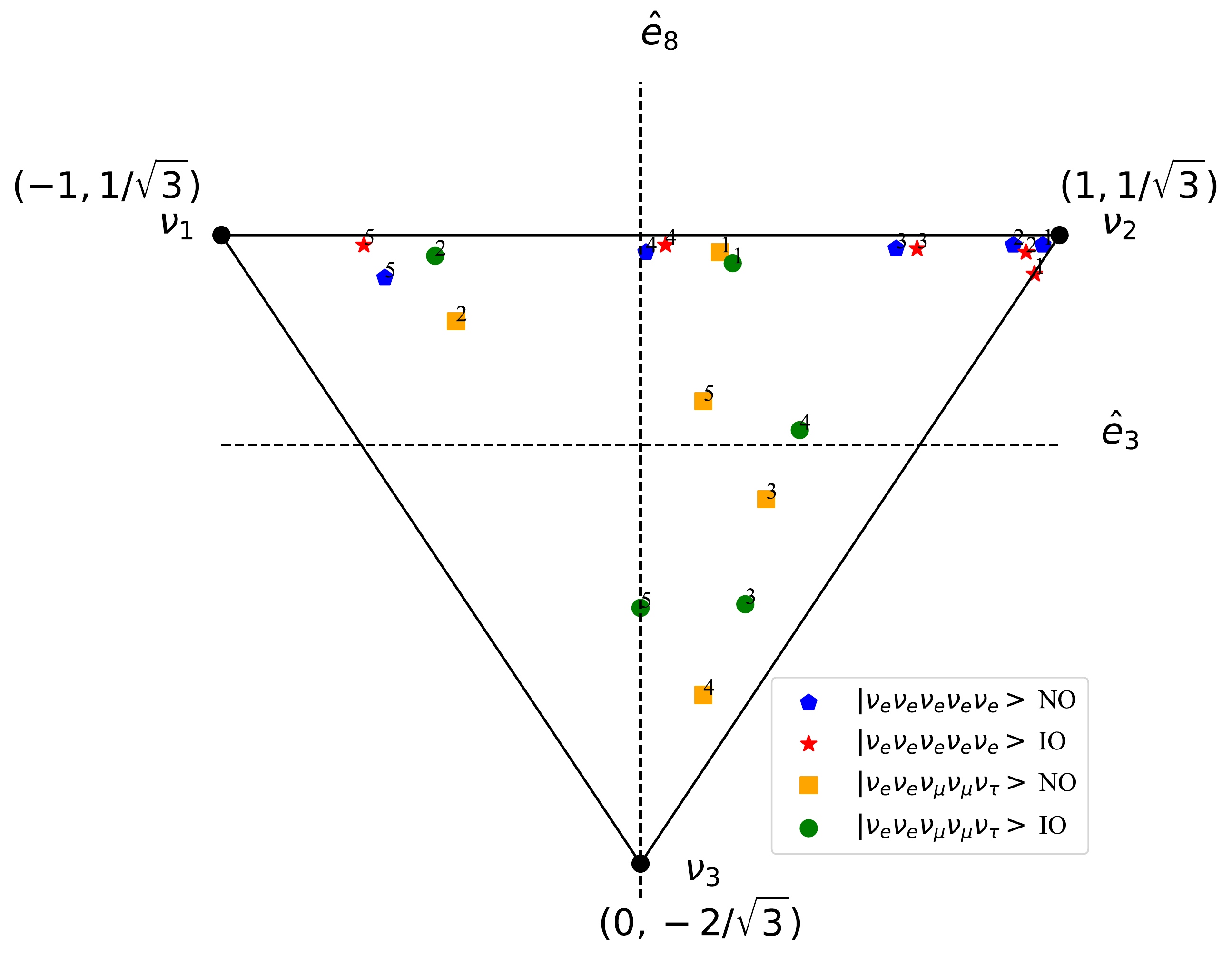}
    \caption{The Asymptotic $P_{3}$ and $P_{8}$ values in $\hat{e}_{3}-\hat{e}_{8}$ plane for $N=3$ (left panel) and $N=5$ (right panel) neutrinos with pure and mixed initial states in both (NO) and (IO). The points are labeled by the neutrino frequency mode, {\it i.e.}, $q$.}
    \label{fig:N3tri}
\end{figure*}

To further elaborate on the behavior of entanglement and mixing of different mass eigenstates in terms of polarization vector components $P_{3}$ and $P_{8}$, we show their values in the asymptotic limits in $\hat{e}_{3}$-$\hat{e}_{8}$ plane in \figureautorefname~\ref{fig:N3tri}. We consider the same cases of initial states as discussed in earlier sections for the systems of three and five neutrinos. The pure mass eigenstates (see \tableautorefname~\ref{tab:p38}) lie on the vertices on an equilateral triangle, and the maximally entangled states are the ones at the centroid of the triangle.

We note that in the case of an initial state where all three neutrinos are in electron flavor, the data points remain close to the $\nu_{1}$-$\nu_{2}$ edge of the triangle. The $P_{8}$ values are close to the extremum {\it i.e.}, $1/\sqrt{3}$ which signifies that third mass eigenstate does not mix significantly with other two. This feature can be attributed to the small value of the mixing angle $\theta_{13}$. Also, due to predominant mixing of only two mass eigenstates, the maximum entropy nearly reaches the limit for a two-flavor system {\it i.e.}, $\sim\log 2$. 
In the IO, as the neutrino with frequency mode $q=1$ is positioned more towards $\nu_{3}$ vertex, it has a larger fraction of the third mass eigenstate as compared to the other neutrinos. In the case of NO, however, it is the neutrino with the maximum frequency mode that has the maximum contribution of third mass eigenstate among all neutrinos. In addition, the first and second mass eigenstates are significantly mixed what leads to larger entanglement as compared to the two flavor case (see left panels of \figureautorefname~\ref{fig:N5} also).

In the case of mixed initial states, there is a significant mixing of the third mass eigenstate as well. As mentioned above, the closer the data point is to the centroid of triangle the more entangled it is. For the three neutrinos system in the NO, the neutrino with frequency mode $q=3$ is closest to the centroid and hence the most entangled one as can also be seen from the entropies shown in \figureautorefname~\ref{fig:N3}. Similarly, in the IO, the neutrinos with second $(q=2)$ and third $(q=3)$ frequency modes are almost equidistant from the centroid, and hence there entropies are almost equal. Furthermore, the data points corresponding to the neutrinos with frequency mode $q=1$ in both mass orderings are closer to the $\nu_{1}$-$\nu_{2}$ edge of the triangle and hence the contribution of the third mass eigenstate is negligible. Similar observations can be made for $q=2$ and $q=3$ mode neutrinos. 

For the five neutrinos case, in NO the neutrino in frequency mode $q=5$ is the the one closest to the centroid followed by neutrino in $q=3$ mode. The neutrino in $q=4$ mode is the closest to one of the vertices {\it i.e.}, $\nu_{3}$, and hence it is the least entangled one with a maximum contribution of third mass eigenstate. The entanglement entropies shown in \figureautorefname~\ref{fig:N5_mix} also follow this ordering. In the case of IO, the neutrino in $q=4$ frequency mode is closest to the centroid followed by the one in $q=5$ mode. The neutrino in $q=2$ frequency mode is the closest to one of the vertices, and hence is the least entangled. These arguments are also supported by the entropies shown in \figureautorefname~\ref{fig:N5_mix}. Therefore, the information obtained from the triangles in  $\hat{e}_{3}$-$\hat{e}_{8}$ plane complements the observations from entropies and the probabilities of finding neutrinos in different mass eigenstates.

\section{Conclusions}
\label{sec:Conclusions}

The collective neutrino oscillations considering the three flavor scenario are investigated in many-body picture for the first time. We have quantified the entanglement in terms of entropies and the conserved polarization vector components. The entanglement is found to be significantly larger, at least in the cases we considered in the present work, as compared to a two-flavor case. Our results thus hint that the entanglement in the neutrino systems may be underestimated in the two-flavor approximation. The results presented here also indicate a greater deviation from the mean-field approximation results (because of large entropy) in which quantum correlations are ignored.

For a small system of five neutrinos and a simple initial state with all particles initially in the electron flavor, the results in the cases of two-flavor and three-flavor evolution show considerable differences.  We have found that the entanglement is substantially larger even for  such a small system with pure initial state, as compared to the two-flavor case, and of course the mean-field approximation. These deviations are expected to be further enhanced for the larger systems and for more complex mixed flavor initial states. We have investigated the impact coming from considering the latter. Our results indicate that the entanglement increases for the mixed initial state as compared to the pure one. In addition, the entanglement also increases with the number of neutrinos considered in the system. These findings are in agreement with what has been found in the two-flavor approximation~\cite{cervia:2019}.

We have also investigated the mixing of different mass eigenstates. The probabilities of finding neutrinos in different mass eigenstates depend strongly on the mass ordering. Furthermore, we have demonstrated the mixing of different mass eigenstate and entanglement by plotting the asymptotic values of conserved flavor-lepton numbers $P_{3}$ and $P_{8}$ in $\hat{e}_{3}$-$\hat{e}_{8}$ plane. This pictorial representation provides complete information on entanglement and mixing of different mass eigenstates, and summarize the information obtained separately from all quantities {\it viz.}, $P_{3}, P_{8}, S, P_{\nu_{1}},P_{\nu_{2}}$, and $P_{\nu_{3}}$.

This work is a first step towards the many-body treatment of collective neutrino oscillations in three flavor settings, and it is not complete in any sense. One can take several directions to explore further in the near future. We list some of them here.

One interesting feature noticed in three-flavor scenario within the mean-field approach is the emergence of multiple spectral splits~\cite{Fogli:2008pt, Fogli:2008fj, Dasgupta:2009mg, Fogli:2009rd, Dasgupta:2010cd}. To investigate the spectral splits, one has to study the evolution of a system with large number of neutrinos.
However, the exponential increase in the size of Hilbert space as $3^{N}$ (as compared to $2^{N}$ in two-flavor case) with the number of neutrinos $N$ in the system, makes these computations complex and computationally expensive. Therefore, with the present numerical approach, we were limited in our calculations to a small number of neutrinos. We note that the use of tensor network techniques, such as ones employed in Ref.~\cite{Cervia:2022pro} for the two-flavor approximation, may allow to investigate the behavior of the spectral splits in the neutrino spectra in case of many-body three flavor treatment of neutrino evolution.

As mentioned in \sectionautorefname~\ref{sec:introduction}, a three flavor neutrino can be considered as a qutrit and therefore the qutrit-based quantum computers, which are expected to be more powerful than the qubit-based one, can be utilized to simulate this system. Therefore, in the future, the quantum information studies based on qutrits can be employed to get further insights into the entanglement in collective neutrino oscillations.

\begin{acknowledgments}

We are grateful for helpful discussions with A~.V.~Patwardhan.
This work was supported in part by the U.S.~Department of Energy, Office of Science, Office of High Energy Physics, under Award  No.~DE-SC0019465 and in part 
by the National Science Foundation Grants No. PHY-1806368, PHY-2020275, and PHY-2108339. A.M.S. and A.B.B would like to thank Kavli Institute for Theoretical Physics for the hospitality during this work. This research was supported in part by the National Science Foundation under Grant No. NSF PHY-1748958. A.M.S., acknowledge a partial support from the Institute for Nuclear Theory at the University of Washington for its kind hospitality and stimulating research environment. This research was supported in part by the INT's U.S. Department of Energy grant No. DE-FG02-00ER41132.

\end{acknowledgments}


\appendix
\section{Closed form for the 3-level entropy}
\label{app:entropy}

In this appendix we present the exact solution to the bipartitie entropy, given in Eq.~\eqref{eq:bipartite-entropy}, for the three level system in terms of the two invariant quantities the magnitude of the polarization vector $|\vec{P}|$ and the $\Pi$ invariant, defined below. 
First we note that the trace of the matrix $\mathcal{A} = \lambda_jP_j$ is
\begin{equation}\label{eq:TrA}
    \mathrm{Tr}\{\mathcal{A}\} = 0 \ ,
\end{equation}
and by using the Gell-Mann matrix identity given by Eq.~\eqref{eq:identity} we can express
\begin{equation}\label{eq:TrA2}
    \mathrm{Tr}\{\mathcal{A}^2\} = \mathrm{Tr}\{\lambda_j \lambda_i P_i P_j  \} = \frac{2}{3}\mathrm{Tr}\{ \delta_{ij} P_i P_j  \mathbb{I} \} = 2|\vec{P}|^2 \ ,
\end{equation}
and 
\begin{equation}\label{eq:TrA3}
    \mathrm{Tr}\{\mathcal{A}^3\} = \mathrm{Tr}\{\lambda_j \lambda_i \lambda_k P_i P_j P_k  \} = 2 d_{ijk} P_i P_j P_k = 2 \Pi \ .
\end{equation}
In order to find the eigenvalues of the matrix $\mathcal{A}$ we solve the characteristic equation for the $3\times3$ matrix, which using Eqs.~\eqref{eq:TrA}-\eqref{eq:TrA3} can be written as
\begin{equation}\label{eq:characteristiction-3x3}
    x^{3} - |\vec{P}|^{2}x - \frac{2}{3} \Pi = 0 \ .
\end{equation}
The roots of this equation, {\it i.e.}, eigenvalues ($x_j$) of matrix $\mathcal{A}$ are given by
\begin{subequations}
\label{eq:eigenvalues} 
\begin{eqnarray}
x_1 &=& \frac{2 |\vec{P}|}{\sqrt{3}} \left[ -\frac{1}{2}\cos \left(\frac{\chi}{3}\right) -\frac{\sqrt3}{2} \sin \left(\frac{\chi}{3}\right) \right] \ , 
\label{eq:x1} \\
x_2 &=& \frac{2 |\vec{P}|}{\sqrt{3}} \left[ -\frac{1}{2}\cos \left(\frac{\chi}{3}\right) +\frac{\sqrt3}{2} \sin \left(\frac{\chi}{3}\right) \right] \ ,
\label{eq:x_2} \\
x_3 &=& \frac{2 |\vec{P}|}{\sqrt{3}} \cos\left(\frac{\chi}{3}\right) \ ,  
\label{eq:x3}
\end{eqnarray}
\end{subequations}
where 
\begin{equation}
\cos\left(\chi\right) = \frac{\sqrt{3}\> \Pi}{|\vec{P}|^3}.
\end{equation}
In addition all the eigenvalues of the density matrix ($\rho$) need to be non-negative. For a $3 \times 3$ matrix this requires the determinant to be non-negative or
\begin{equation}
1- 3 \> \mathrm{Tr} \rho^2 + 2 \> \mathrm{Tr} \rho^3 \ge 0
\end{equation}
or
\begin{equation}
\Pi \ge |\vec{P}|^2 -\frac{4}{9}
\end{equation}
where the equality holds only for a pure state.

We can rewrite Eq.~\eqref{eq:entropy-Baha} using the eigenvalues of the matrix $\mathcal{A}$ as
\begin{equation}\label{eq:entropy-app}
    S_n = \log 3  - \frac{1}{3} \sum_j \left(1+\frac{3}{2}x_j\right)\log\left(1 + \frac{3}{2}x_j\right)\ ,
\end{equation}
where the first part of the sum using the Eq.~\eqref{eq:eigenvalues} simplifies to 
\begin{equation}\label{eq:first-part-of-the-sum}
    \sum_j \log\left(1+\frac{3}{2}x_j\right) = \log\left(1 - \frac{9}{4} \left(|P|^2  - \Pi \right)\right) \ .
\end{equation}
The second term in the sum from Eq.~\eqref{eq:entropy-app} has a more convoluted form of
\begin{eqnarray}
 \label{eq1}
\begin{split}
 \sum_j \frac{3}{2}x_j\log\left(1+\frac{3}{2}x_j\right)  =& \; \frac{3}{2} ac \log\left(\frac{(\frac{2}{3}+x_1)(x_2+\frac{2}{3})}{(x_3+\frac{2}{3})^2}\right) \\ 
 + & \; \frac{3}{2} bc \log\left(\frac{(x_2+\frac{2}{3})}{(x_1+\frac{2}{3})}\right) \ ,
\end{split}
\end{eqnarray}
where the coefficients $a$, $b$, and $c$ are
\begin{subequations}
\label{eq:constants-app} 
\begin{eqnarray}
a &=&  -\frac{1}{2} \cos \left(\frac{\chi}{3}\right) \ , 
\label{eq:A} \\
b &=& \frac{\sqrt{3}}{2}  \sin \left(\frac{\chi}{3}\right) \ ,
\label{eq:B} \\
c &=& \frac{2 |\vec{P}|}{\sqrt{3}} \ .
\label{eq:C}
\end{eqnarray}
\end{subequations}
Plugging-in the above the expressions into the Eq.~\eqref{eq1} we get 
\begin{equation} \label{eq:second-part-of-the-sum}
\begin{split}
 & \frac{3}{2} \sum_j x_j \log \left( 1 + \frac{3}{2} x_j \right)   \\
& = \frac{3}{2} |\vec{P}|\sin \left(\frac{\chi}{3}\right) \log\left(\frac{1-\sqrt{3}|P|\cos((\pi + \chi)/3)}{1-\sqrt{3}|P|\cos((\pi- \chi)/3)} \right) \\
 & + \frac{3\sqrt{3}}{2}|\vec{P}|\cos \left(\frac{\chi}{3}\right) \log 
 \left(1 -\frac{3}{4}\frac{ \left(3 \vec{|P|}^2-4\right)}{ \left(1 + \sqrt{3} \vec{|P|} \cos \left(\frac{\chi}{3}\right)\right)^2}\right.\\
 &\left.-\frac{3}{1 + \sqrt{3} \vec{|P|} \cos \left(\frac{\chi}{3}\right)} \right)   \ .
 \end{split}
\end{equation}
Substituting the expressions for first and second parts of sum given in Eqs.~\eqref{eq:first-part-of-the-sum} and~\eqref{eq:second-part-of-the-sum}, respectively, in Eq.~\eqref{eq:entropy-app}, we get the closed form of bipartite entanglement entropy of $n$-th neutrino.


\vskip 1cm

\newpage
\phantom{i}
\newpage
\bibliographystyle{apsrev}
\bibliography{3-level}

\end{document}